\documentclass{ws-ijmpd}

\begin{document}

\markboth{ Das and Chakrabarti}
{Properties of Accretion Shocks in Viscous Flows with Cooling Effects}

%
\catchline{}{}{}{}{}
%

\title{Properties of Accretion Shocks in Viscous Flows with Cooling Effects}

\author{Santabrata Das and Sandip K. Chakrabarti\footnote{Also at 
Centre for Space Physics, Chalantika 43, Garia Station Rd., Kolkata 700084}}

\address{S.N. Bose National Centre for Basic Sciences,\\
JD-Block, Sector III, Salt Lake, Kolkata 700098, India\\
sbdas@bose.res.in; chakraba@bose.res.in}

\maketitle

\begin{history}
\received{Day Month Year}
\revised{Day Month Year}
\end{history}

\begin{abstract}
Low angular momentum accretion flows can have standing and oscillating shock waves.
We study the region of the parameter space in which multiple sonic points
occur in viscous flows in presence of various cooling effects such as bremsstrahlung
and Comptonization. We also quantify the parameter space in which shocks are steady or oscillating.
We find that cooling induces effects opposite to heating by viscosity
even in modifying the topology of the solutions, though one can never be exactly balanced
by the other due to their dissimilar dependence on dynamic and thermodynamic parameters.
We show that beyond a critical value of cooling, the flow ceases to contain a shock wave.

\end{abstract}


\noindent Accepted for publication in Int. J. Mod. Phys. (D)

\section{Introduction}	

Cooling and heating processes play an important role in studying the accretion disks around compact objects.
In this paper, we consider both viscous heating
and bremsstrahlung cooling as energy dissipative processes.
In Chakrabarti and Das$^1$, the problem of
accretion and winds with small angular momentum was solved
in a very comprehensive way when viscous heating is included.
The entire parameter space was scanned and separated
in terms of flows with and without shocks. A major scope for further
work is to find and separate the parameter space which allows
shock formations when cooling effects are also included. This will allow us to
study flows with high accretion rate as well. The question arises because viscosity
transports angular momentum and increases the possibility of shock
formation at a larger distance from the black hole. However, cooling reduces the
post-shock pressure and therefore the possibility of shock formation. Various
cooling processes will and should change the parameter space in which shocks form.

In Chakrabarti$^2$, some of the effects of cooling was
included. Assuming that cooling at a given radius of an accretion disk
is proportional to the heating, all possible topologies were presented.
While cooling will and should depend on heating, there is no reason to believe that
this proportionality is independent of radial distance. It is therefore
necessary to re-visit the problem with explicit form of heating and cooling included.

In the present Paper, we make these important extensions of the
previous work and show that shocks are still possible in a very large
part of the parameter space. We find topologies of solutions which
are similar to what was found for spiral shock study$^3$, but
otherwise new for axisymmetric situation. We also find new topologies which were
not anticipated before. In more recent years, it has become evident that
the standing shocks may be very important in explaining the spectral properties of
black hole candidates$^4$ as the
post-shock region behaves as the boundary layer where accreting
matter dissipates its thermal energy and generates hard X-ray
by inverse Comptonization. This region is also found to be responsible to
produce relativistic outflows. Furthermore, numerical simulations indicated that
the shocks may be oscillating at nearby regions of the parameter space
in presence of cooling effects$^5$
and the shock oscillations can also explain intricate properties
of quasi-periodic oscillations$^6$. 
Recent observations do support the presence of sub-Keplerian
flows in accretion disks$^{7-8}$.

The present work is  done
around a Schwarzschild black hole by using pseudo-Newtonian potential$^9$.
We use a similar viscosity prescription as in Chakrabarti$^2$.
In \S 2, we present model equations which included  both heating and cooling effects.
In \S 3, we perform the sonic point analysis. In \S 5, we study the nature of the sonic
points and how they vary with flow parameters.
In \S 6, we study the global solution topologies with heating and cooling effects.
In \S 7, we classify the parameter space in terms of whether shocks will form or not
and how it depends on flow parameters.  In \S 8, we briefly report how even the super-Keplerian
flows may also be transonic. Finally,
in \S 9, we discuss the importance  of these solutions and make concluding remarks.

\section {Model Equations When Cooling Effects are Included}

As far as the cooling processes are concerned,
they could be due to various physical reasons, such as
the thermal and the non-thermal bremsstrahlung, synchrotron,
Comptonization etc.  For simplicity, we assume that the
Comptonization enhances the injected photon intensity due to bremsstrahlung
by a factor of $\zeta$ which can take any value from $1$ to
$\sim$  few $ \times 100$ depending on the availability of soft photons$^4$. In other words, we use $\zeta$ as a
parameter to represent the net cooling.

We start with a thin, axisymmetric, rotating viscous accretion flow around a Schwarzschild
black hole. The space-time geometry around a non-rotating black hole can be satisfactorily  described by
the pseudo-Newtonian potential$^9$ and is given by $ g(x)=-\frac{1}{2(x-1)}$, 
where, $x$ is the radial distance in dimensionless unit.
In the steady state, the dimensionless hydrodynamic equations
that govern the infalling matter are the followings$^2$.

\noindent (a) Radial momentum equation :

$$
\vartheta \frac {d\vartheta}{dx}+\frac {1}{\rho}\frac {dP}{dx}
-\frac {\lambda(x)^2}{x^3}+\frac {1}{2(x-1)^2}=0 .
\eqno{(1a)}
$$

\noindent (b) Baryon number conservation equation :

$$
\dot M = \Sigma \vartheta x ,
\eqno{(1b)}
$$
apart from the geometric constant.

\noindent (c) Angular momentum conservation equation :
$$
\vartheta \frac {d\lambda(x)}{dx}+\frac{1}{\Sigma x}
\frac {d}{dx}\left( x^2 W_{x\phi}\right)=0
\eqno{(1c)}
$$

\noindent (d) The entropy generation equation :

$$
\Sigma \vartheta T \frac {ds}{dx}= Q^+ - Q^-
\eqno{(1d)}
$$
Here, $\vartheta$ is the radial velocity and $\lambda(x)$  
is the specific angular momentum of the flow. The distances, velocities and masses
are made dimensionless by using $r_g=2GM_{BH}/c^2$, the Schwarzschild radius,
$c$, the velocity of light and $M_{BH}$, the mass of the black hole, 
respectively. Here $\Sigma$ and $W_{x\phi}$ are the
vertically integrated density and viscous
stress, $s$ is the entropy density of the flow, $T$ is the local temperature,
$Q^+$ and $Q^-$ are the heat gained and lost by the flow, and $\dot M $ is the mass
accretion rate. We assume that the accretion flow is in hydrostatic
equilibrium in the vertical direction and the vertical velocity component is much
smaller compared to the radial component. With this assumption, the local 
disk height is obtained by equating the pressure gradient force in the vertical 
direction with the gravitational force. The half thickness of the disk is then given by,
$$
h(x)=ax^{1/2}(x-1),
\eqno{(2)}
$$
where, the sound speed is defined as $a=\sqrt {\gamma P/\rho}$, where $\gamma$, $P$ and 
$\rho$ being the adiabatic index, pressure and density respectively.

In the present paper, we follow a similar viscosity prescription as given in
Chakrabarti$^2$ where $W_{x\phi}=-\alpha_{\Pi}\Pi$ is used. This 
prescription ensures that viscous stress remain continuous at the flow 
discontinuity (shock) in presence of significant radial motion of the accretion 
flow.

\section{Sonic Point Analysis}

A black hole accretes matter either from its binary companion or
from the surrounding ambient medium. This matter starts with a negligible
radial velocity at the outer edge of the disk. But it enters into the 
black hole with a velocity of light$^{10}$. This ensures that inside the accretion 
disk there must be  at least one point where the radial velocity
exactly matches with the sound speed. This point is known as
the sonic point.
Accretion flow which passes through a shock wave the must 
cross sonic points at least twice. In other words, the flow may be called
multi-transonic.  

For the accretion flow, entropy equation (1d) can be simplified as,
$$
\frac {\vartheta}{\gamma -1}\left[ \frac {1}{\rho}
\frac {dP}{dx}-\frac {\gamma P}{\rho^2}\frac {d\rho}{dx}\right]
=\frac {Q^--Q^+}{\rho h}=C-H .
\eqno{(3)}
$$
and then $H (= Q^+/\rho h)$ takes the form,
$$
H = A x (ga^2+\gamma\vartheta^2)\frac {d\Omega}{dx} 
\eqno{(4)}
$$
where, 
$A = -\alpha_{\Pi}\frac {I_n}{\gamma}$ and $g = \frac
{I_{n+1}}{I_n}$. Here, $\Omega(x)$ is the angular velocity of the
accreting matter at the radial distance $x$, $n$ is the polytropic index
$(n= \frac {1}{(\gamma-1)})$, $I_n$ and $I_{n+1}$ come
form the vertically averaged density and pressure.
The general expression of $I_n$ is given by$^{11}$:
$$
I_n = \frac {(2^n n!)^2}{(2n + 1)!} .
\eqno{(5)}
$$

Simultaneously with viscous heating, we use Comptonization of the bremsstrahlung
radiation as the physical cooling process. 
The following analysis is carried out with non-dimensional cooling term $C (=Q^-/\rho h)$ as
$$
C = \frac {\zeta B}{\vartheta x^{3/2}(x-1)},
\eqno{(6a)}
$$
with
$$
B = 1.4 \times 10^{-27} \left( \frac {\mu m_p}{2k}\right)^{1/2}
\frac {\dot M}{2\pi m^2_p}\frac {1}{2GcM_{BH}} ,
\eqno{(6b)}
$$
$\mu$ is the mean molecular weight, $m_p$ is the mass of the proton and $k$ is the Boltzmann
constant respectively.

\subsection{Sonic Point Condition}

From equations (1a-1d), the sonic point conditions are derived following the general procedure$^{12}$
and are given by,

$$
\frac {d\vartheta}{dx}=\frac {N}{D},
\eqno{(7)}
$$
where the numerator $N$ is
$$
N =-\frac {\alpha_\Pi A (a^2g+\gamma \vartheta^2)^2}{\gamma x}
-\left[ \frac {\lambda^2}{x^3}-\frac {1}{2(x-1)^2}\right]
\left[ 2\alpha_\Pi g A (a^2g+\gamma\vartheta^2)+ \frac {(\gamma+1)
\vartheta^2}{(\gamma-1)} \right]
$$
$$
-\frac {\vartheta^2 a^2(5x-3)}{x(\gamma-1)(x-1)}
-\frac {\alpha_\Pi g A a^2(5x-3)(a^2g+\gamma\vartheta^2)}{\gamma x(x-1)}
+\frac {2\lambda A \vartheta (a^2g+\gamma\vartheta^2)}{x^2}
+\frac {B}{x^{3/2}(x-1)}
\eqno{(8a)}
$$
and the denominator $D$ is,
$$
D = \frac {2a^2\vartheta}{(\gamma-1)}-\frac {(\gamma+1)\vartheta^3}
{(\gamma-1)}
-A\alpha_\pi\vartheta(a^2g+\gamma\vartheta^2)
\left[ (2g-1)-\frac {a^2g}{\gamma\vartheta^2}\right] .
\eqno{(8b)}
$$

At the sonic point, both the numerator and denominator vanishes. For
$D=0$, one can get the expression for the Mach Number $M(x_c)$ at the sonic
point which is given by,
$$
M(x_c) =\sqrt {\frac{-m_b - \sqrt{m^2_b-4m_a m_c}}{2m_a}}
\approx \sqrt {\frac {2}{\gamma +1}} \ \ {\rm for} \ \
\alpha_{\Pi} \rightarrow 0 ,
\eqno{(9)}
$$
where,
$$
m_a=-A\alpha_\Pi \gamma^2(\gamma-1)(2g-1)-\gamma(\gamma+1),
\eqno{(10a)}
$$
$$
m_b=2\gamma-2A\alpha_\pi g\gamma(\gamma-1)(g-1) ,
\eqno{(10b)}
$$
$$
m_c=A\alpha_\pi g^2 (\gamma-1).
\eqno{(10c)}
$$

In order that the Mach number to be physically acceptable,
we do not use the negative sign within the square root.
In the weak viscosity limit, Mach number at the sonic point reduces to the result as
obtained in Chakrabarti$^{12}$.

Setting $N = 0$, we get an algebraic equation for sound speed at the
sonic point which is given by,
$$
F({\cal E}_c, \lambda_c, x_c)={\cal A}a^4(x) + {\cal B}a^3(x)
+ {\cal C}a^2(x) +{\cal D} = 0 ,
\eqno{(11)}
$$
where,
$$
{\cal A} = -\left[ \frac {\alpha_\Pi A \{g+\gamma M^2\}^2}
{\gamma x}+\frac {\alpha_\Pi A (5x-3)\{g+\gamma M^2\}}
{\gamma x(x-1)}+\frac {M^2(5x-3)}{x(\gamma-1)(x-1)} \right],
\eqno{(12a)}
$$

$$
{\cal B} = \frac {2\lambda A M (g+\gamma M^2)}{x^2},
\eqno{(12b)}
$$
$$
{\cal C} = -\left[ \frac {\lambda^2}{x^3}-\frac {1}{2(x-1)^2}\right]
\left[ 2\alpha_\Pi g A (g+\gamma M^2)+ \frac {(\gamma+1)
M^2}{(\gamma-1)} \right],
\eqno{(12c)}
$$
$$
{\cal D} = \frac {B}{x^{3/2}(x-1)}.
\eqno{(12d)}
$$

We solve the above quadratic equation to obtain the sound speed at the sonic point.
Das, Chattopadhyay and Chakrabarti$^{13}$ found that depending on a given set of initial parameters
accretion flow may have a maximum of four sonic points where one of the sonic points always lies
inside the black hole horizon for non-dissipative accretion flow$^{10, 13}$.
In our present study we also expect the similar result if we are interested in the weak viscosity 
and weak cooling limit. Below, we study the nature of the sonic points.

\section{Nature of the sonic points}

Accreting matter begins subsonically from the outer edge of the disk and becomes 
supersonic after passing through the sonic point before entering into black hole.
A flow may contain multiple sonic points depending on the initial set of 
input parameters. Nature of sonic points depends on the value of the velocity
gradients $(d\vartheta/dx)$ at the sonic points. At each sonic point 
$(d\vartheta/dx)$ has distinctly two different values.
If both the velocity gradients are real and opposite   
signs, the sonic points is saddle type and one is used for accretion and the 
other used for winds. Nodal type sonic point belongs to the
class when  the derivatives are real and of the same sign. When the derivatives are complex, the 
sonic point is of spiral type. For a standing shock to form, an accretion flow must have
more than one saddle type sonic points. 

\begin {figure}
\vbox{
\centerline{
\psfig{figure=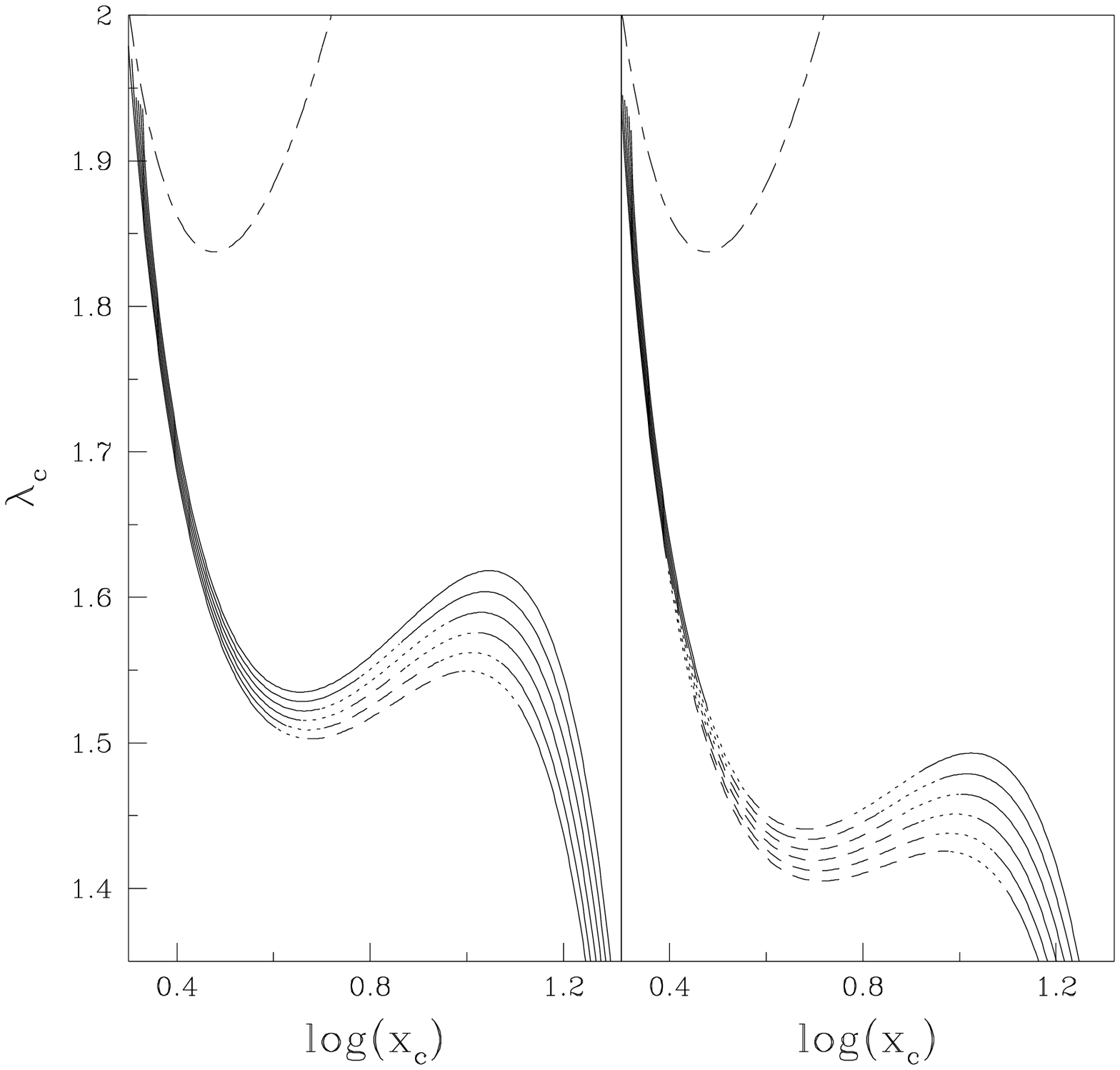,height=10truecm,width=12truecm}} }
\noindent{\small {\bf Fig. 1(a-b):} Variation of specific angular momentum ($\lambda_{c}$)
as a function of the logarithmic sonic point location $(x_c)$ for the viscosity
parameter (a) $\alpha_\Pi = 0.1$ (left panel) and (b) $\alpha_\Pi = 0.5$ (right panel). 
Dimensionless accretion rate ${\dot m}=1.0$ and energy at the sonic point ${\cal E}_c=0.0013$ are
chosen. Long-dashed-dotted curve in the upper part is the Keplerian
angular momentum distribution. Solid curves represent the saddle type sonic points, 
dotted curves represent the nodal type sonic points and the short-dashed curves are 
for the spiral type sonic points. The Comptonization cooling factor $\zeta$ is
$1$ (bottom curve), $20$, $40$, $60$, $80$ and $100$ (top curve) respectively. Clearly, higher
cooling and higher viscosity remove the outer sonics. Eventually the disk becomes a 
Keplerian disk passing through the inner sonic point.}
\end{figure}

In Figure 1a, we plot the variation of specific angular momentum ($\lambda_{c}$) 
as a function of the logarithmic sonic point location $(x_c)$ for a given viscosity
parameter ($\alpha_\Pi = 0.1$), and dimensionless accretion rate (made dimensionless by the Eddington rate)
${\dot m}=1.0$ and a given specific energy ${\cal E}_c = 0.013$
at the sonic points. The variable used in cooling efficiency factor $\zeta$ (Eqn. 6a).
From the bottom curve to the top, $\zeta= 1,\ 20,\ 40,\ 60,\ 80$ and $100$ respectively. 
In Fig. 1b, the same curve is shown for $\alpha_\Pi=0.5$, other parameters remaining the same. 
The long-dashed curve at the top represents the Keplerian   
angular momentum distribution which is completely independent of the initial flow
parameters and depends only on the geometry. Solid part of the curves
represents the saddle type sonic points, dotted curve represents the nodal type sonic points
and the short-dashed curves are for the spiral type sonic points. 
Here, at a higher viscosity, the number of sonic points 
becomes three even with very low angular momentum. For no Comptonization (lowermost
curve), the viscous heating is so strong that only outermost sonic point (solid part of the curve
at large radius) exists. Only a large degree of cooling can compensate for the viscous heating 
to bring back the innermost sonic point.

First notice that the sonic points occur at angular momentum below
Keplerian value. For lower values of cooling at the sonic point, an accretion flow contains
all the three types of sonic points in a systematic order: saddle --- nodal ---
spiral --- nodal --- saddle for monotonic increase of location of sonic points.
With the increase of $\zeta$ the region of spiral type sonic points
gradually decreases and finally replaced by the nodal type sonic
points though multiple sonic points still exist.
For further increase of $\zeta$ all the nodal type sonic points also
disappear and replaced by saddle type sonic points. Note that the angular momentum
is always sub-Keplerian. Later, we shall show that
when the cooing is very strong, sonic points will form even for super-Keplerian
flows.

We continue our investigation of the transonic nature of the flow by showing in Fig. 2
a series of curves where the specific energy at the sonic point is changed (marked). The long dashed curve
is the Keplerian distribution as before. Meanings of solid, dashed and dotted curves are
the same as before. For negative energies there are two sonic points, the inner one is 
saddle type (shown in solid curve) and the dashed curve is the spiral type. For each energy, two curves are
drawn. The thick curve is drawn when both the heating and cooling are included while the thin curve is 
drawn when only the heating is included. The motivation is to impress that the character of a
solution can be changed when cooling is included. For instance the solution for
${\cal E}_c=0.019$ with heating and cooling has no spiral or nodal sonic points. But when the
cooling is turned off, the saddle type point becomes nodal type.

\begin {figure}
\vbox{
\centerline{
\psfig{figure=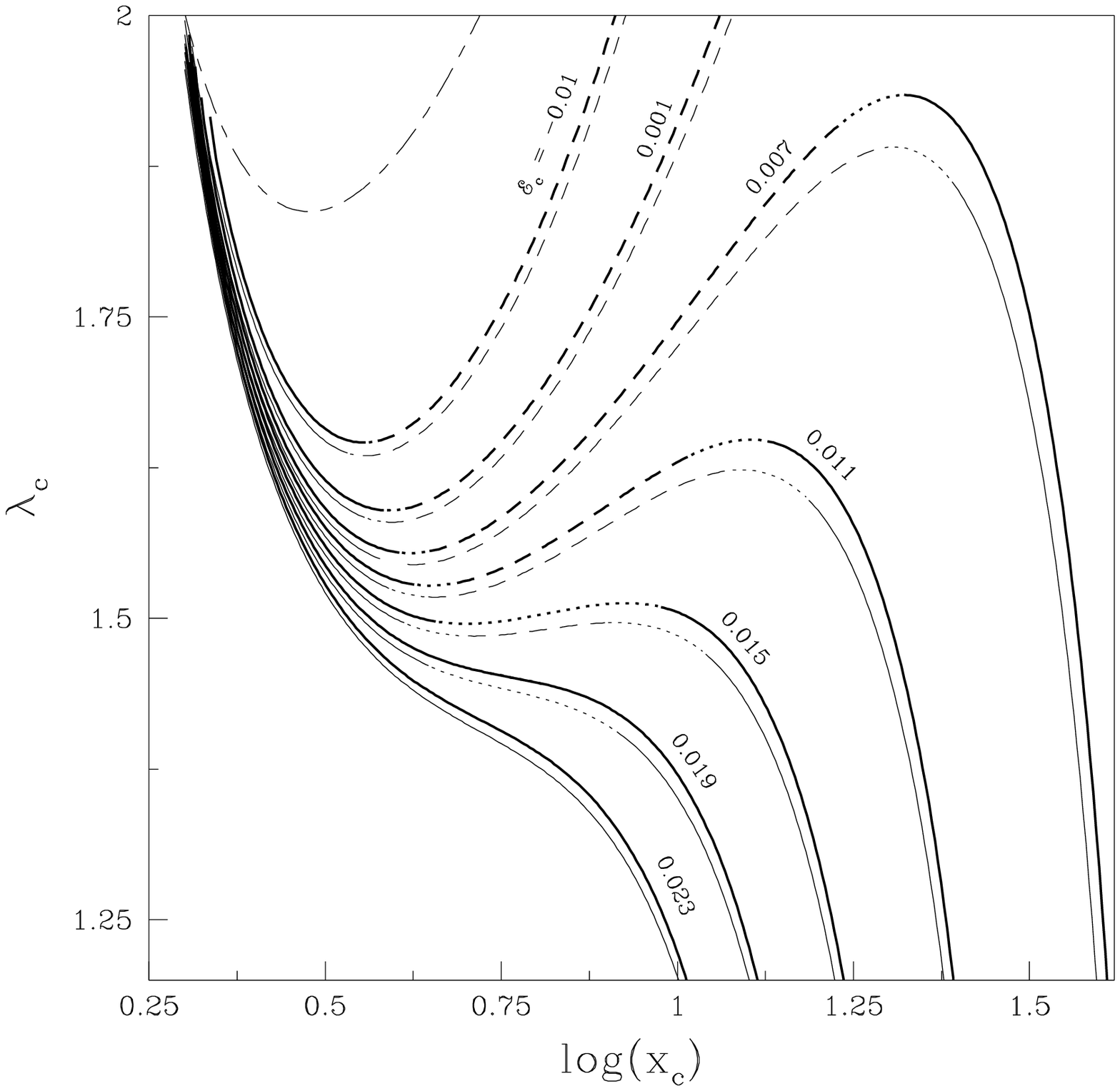,height=10truecm,width=10truecm}
}}
\noindent{\small {\bf Fig. 2:}
Variation of the specific angular momentum $(\lambda_c)$ at the 
sonic point $(x_c)$ as a function of the
specific energy $({\cal E}_c)$ of the flow.  For each energy, two curves are
drawn. The thick curves are drawn for the cases when both the heating and cooling 
are included while the thins curves are drawn when only the heating is 
included. For negative energies there are only two sonic points.}
\end{figure}

In our further study of the nature of the sonic points we draw in Figs. 3(a-b) the 
variation of energy at the sonic points with a cooling factor. In Fig. 3a,
low viscosity ($\alpha_\Pi=0.1$) is used, while in Fig. 3b, high viscosity 
($\alpha_\Pi=0.5$) is used. Other parameters are: $\lambda=1.65$ and ${\dot m}=1$. The curves 
from bottom to the top are for the cooling factor $\zeta= 1, 20, 40, 60, 80, 100$ and $120$.
Notice that three sonic points occur only when the specific energy is positive, i.e.,
for sufficiently hot Keplerian disks or sub-Keplerian flows. For high viscosity outer sonic 
points almost  disappear from regions close to the black hole.

\begin {figure}
\vbox{
\centerline{
\psfig{figure=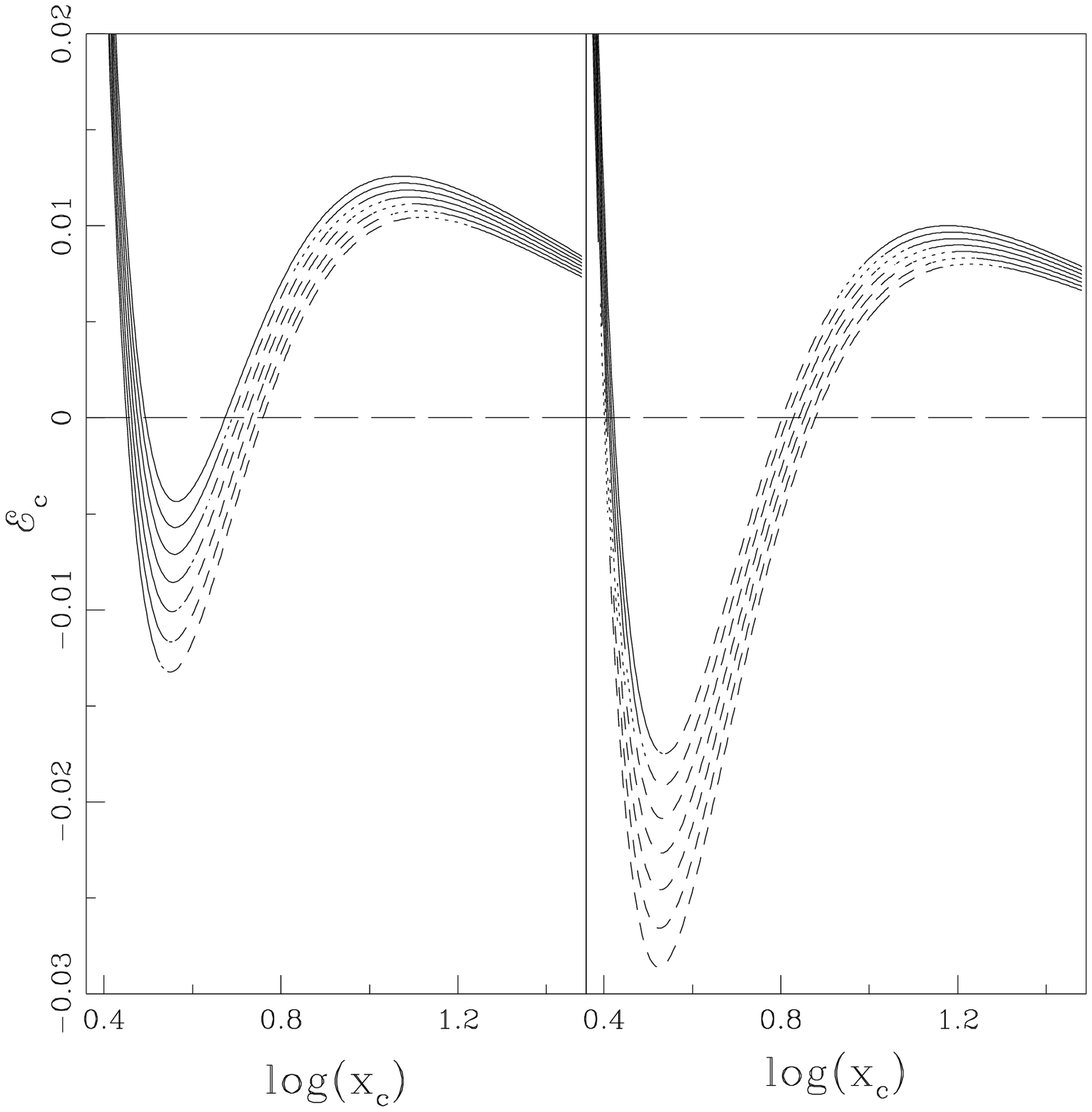,height=10truecm,width=10truecm}
}}
\noindent{\small {\bf Fig. 3(a-b):}
Variation of the specific energy $({\cal E}_c)$ at the sonic point $(x_c)$ as a 
function of the cooling rate of the flow. (a) $\alpha_\Pi=0.1$ (left panel) 
and (b) $\alpha_\Pi=0.5$ (right panel). 
For high viscosity, the outer sonic points almost disappear. }
\end{figure}

In Fig. 4, we show how the number of sonic points, reduced due to viscous heating process,
is recovered back with the introduction of cooling. The curves are drawn, from 
the bottom to the top, for $\zeta=1, \ 20,\  60,\ 80 \ 100$ and $140$ respectively. 
Solid and dotted curves are for saddle and nodal type sonic points respectively.
Other parameters are $\alpha_\Pi=0.1$ and ${\dot m}=1$. Polytropic index $n=1.75$ 
and specific energy ${\cal E}=0.0018$. The long dashed curve 
gives the boundary  of the angular momentum $\lambda_{crit}=1.1733$ and 
cooling factor $\zeta=4.2$ below which there are no triple sonic points, i.e., no shocks in steady flows.

\begin {figure}
\vbox{
\centerline{
\psfig{figure=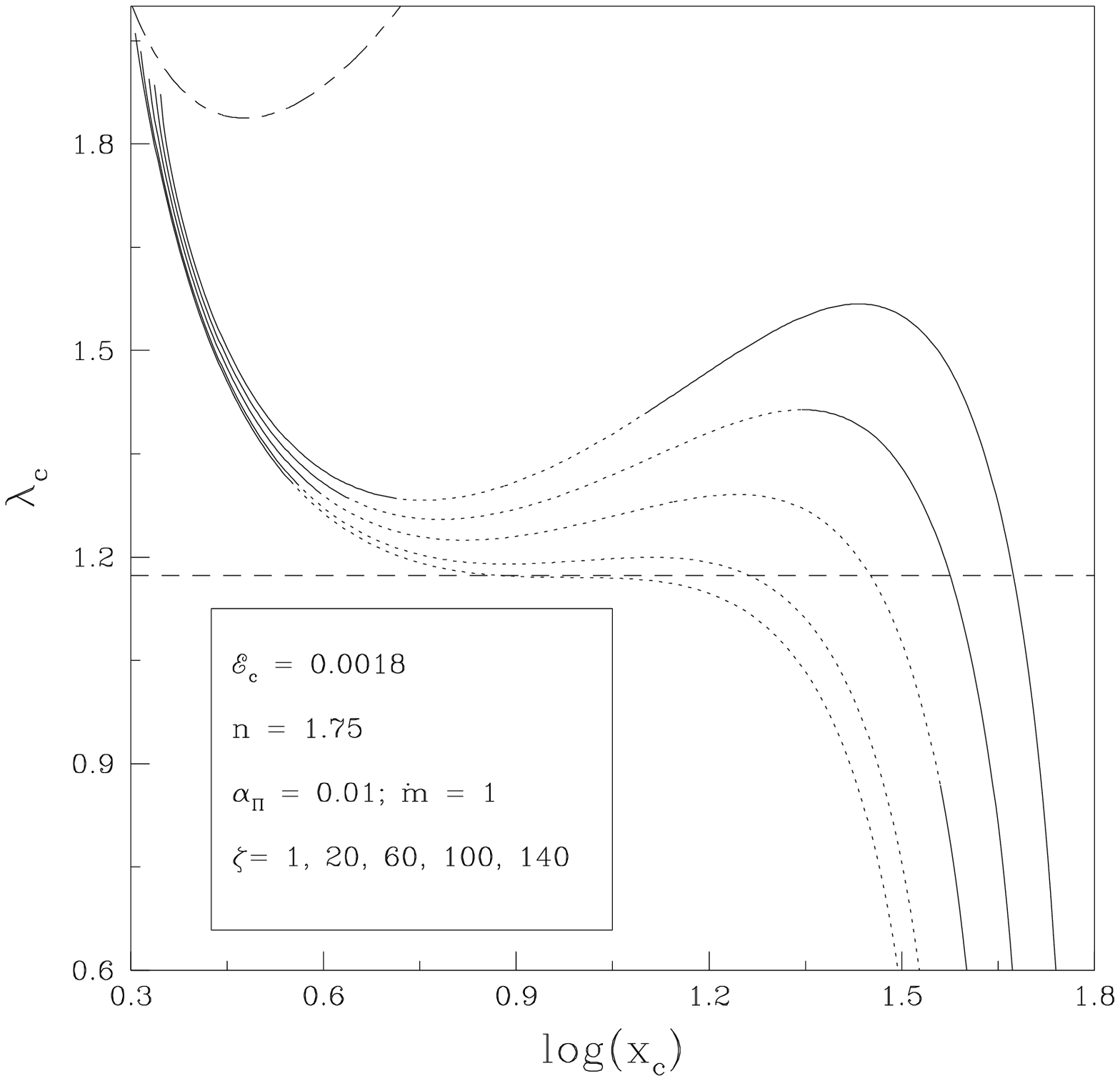,height=10truecm,width=10truecm}
}}
\noindent{\small {\bf Fig. 4:}
Recovery of outer sonic point points as cooling processes are introduced. Solid and
dotted curves are for saddle and nodal type sonic points respectively. Parameters are marked on the plot. Long dashed
curve at $\lambda_c=1.1733$ gives a boundary below which there are no triple sonic points. }
\end{figure}

\section{Global Solution Topology}

Study of shock properties require multi-transonic accretion flow. Accretion flow
passes through two different saddle type sonic points and discontinuous jump of the
flow variable joins these two different branches---one passes through the 
inner sonic point and the other passes through the outer sonic point. 
This discontinuous jump usually known as standing shock transition. In this paper 
we discuss the nature of solution topology in presence of viscous heating and 
bremsstrahlung cooling.

In Fig. 5a, we show how the solution topologies change with cooling. 
Here we plotted the Mach number as a function of the logarithmic radial distance 
for $\zeta=1$, $10$, $25$ and $50$ respectively. Other chosen parameters
are: $x_{in}=2.71$, $\lambda_{in}=1.68$, $\alpha_\Pi=0.01$, ${\dot m}=0.5$ respectively. We note that the
topology opens up to allow flows to enter into black holes through the inner sonic points. 

In Fig. 5b, we show six panels in which we assume higher accretion rate and higher 
inner sonic point locations. The parameters are: $\zeta=1,\ 10, \ 25, \ 33.1, \ 50$ and $70$ respectively.
Other parameters are: $x_{in}=3.5$, $\lambda_{in}=1.68$, $\alpha_\Pi=0.01$, ${\dot m}=0.2$.
Here, as the cooling is increased, the topologies open up similarly, 
but the route to opening up is different. For instance, the 
solution in the fourth panel, with $\zeta=33.1$ is completely new and intriguing. In this case the flow
has the potential to join with a Keplerian disk far away (with low Mach number), while at the same time,
it also passes through the inner sonic point. But it has multi-valued solution: there are
two Mach numbers at a given radial distance in some region. We conjecture that this
type of solution should be unstable and would cause non-steady accretion.

\begin {figure}
\vbox{
\centerline{
\psfig{figure=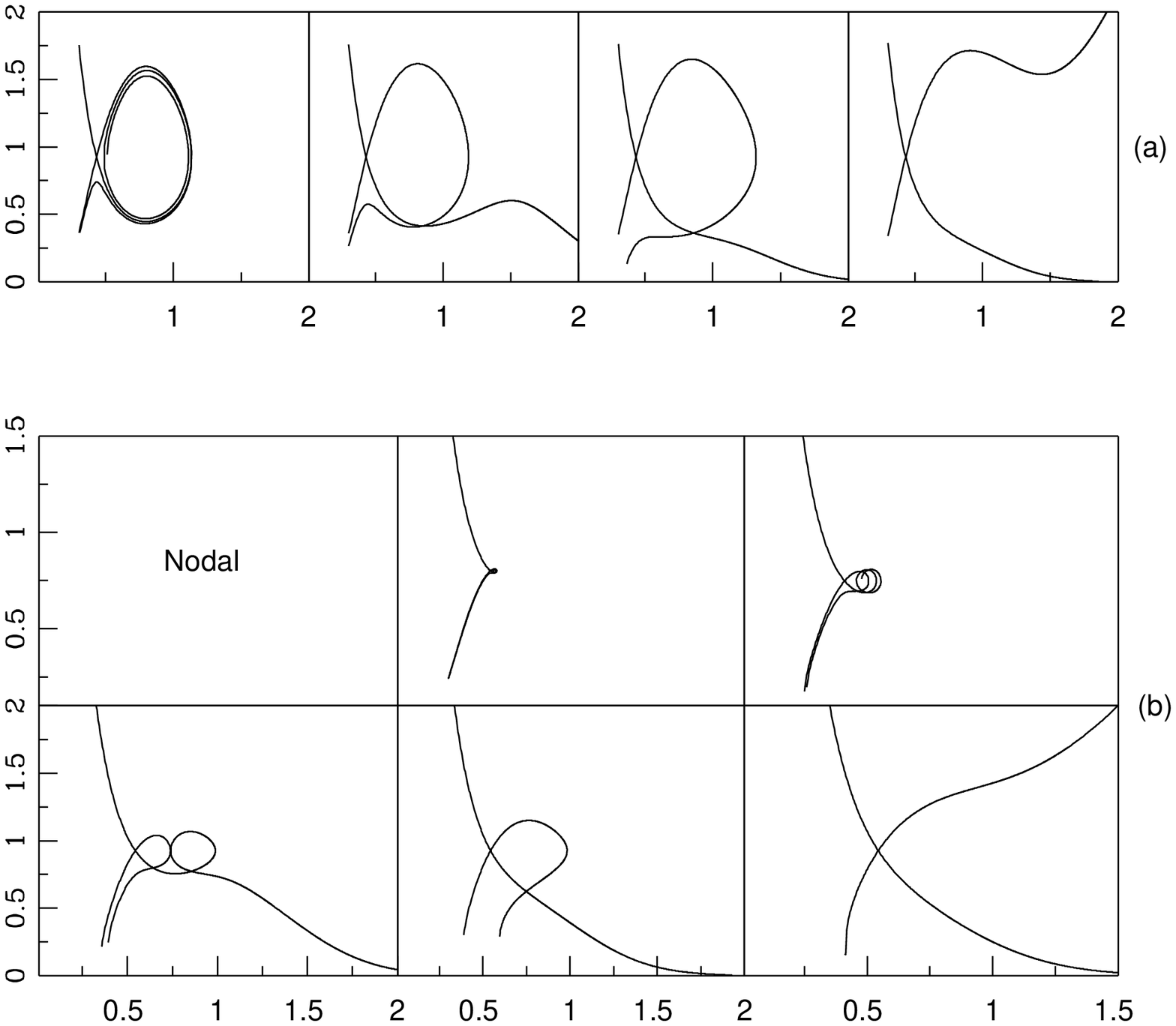,height=10truecm,width=8truecm}
}}
\noindent{\small {\bf Fig. 5:}
Solution topologies in presence of heating and cooling. In (a), the parameters are $\zeta=1, \ 10, \ 25$ and $50$
respectively. 
Other parameters are $x_{in}=2.71$, $\lambda=1.68$, $\alpha=0.01$, ${\dot m}=0.5$.
In (b), the parameters are $\zeta=1,\ 10, \ 25, \ 33.1, \ 50$ and $70$ respectively.
Other parameters are; $x_{in}=3.5$, $\lambda=1.68$, $\alpha=0.01$, ${\dot m}=2.0$. }
\end{figure}

Many of these topologies show `multiple crossings' very similar to what was found in the study of 
spiral shocks$^3$. Actually, since entropy is changing along the flow, the two dimensional
nature of the plots in Fig. 5 is slightly misleading.
In Fig. 6, we show a 3-dimensional plot of the second panel ($\zeta=10$) of Fig. 5a
in which the specific entropy is also plotted as it varies along the flow. In particular, this diagram shows that
at the true sonic point, the specific entropy is identical in both the incoming (solid) and outgoing 
(dotted) branches. But at the `intersection' (at around $M\sim 0.5$) marked by two squares, 
the entropies are completely different. Thus, there is no `sonic point' around $M\sim 0.5$. 

\begin {figure}
\vbox{
\centerline{
\psfig{figure=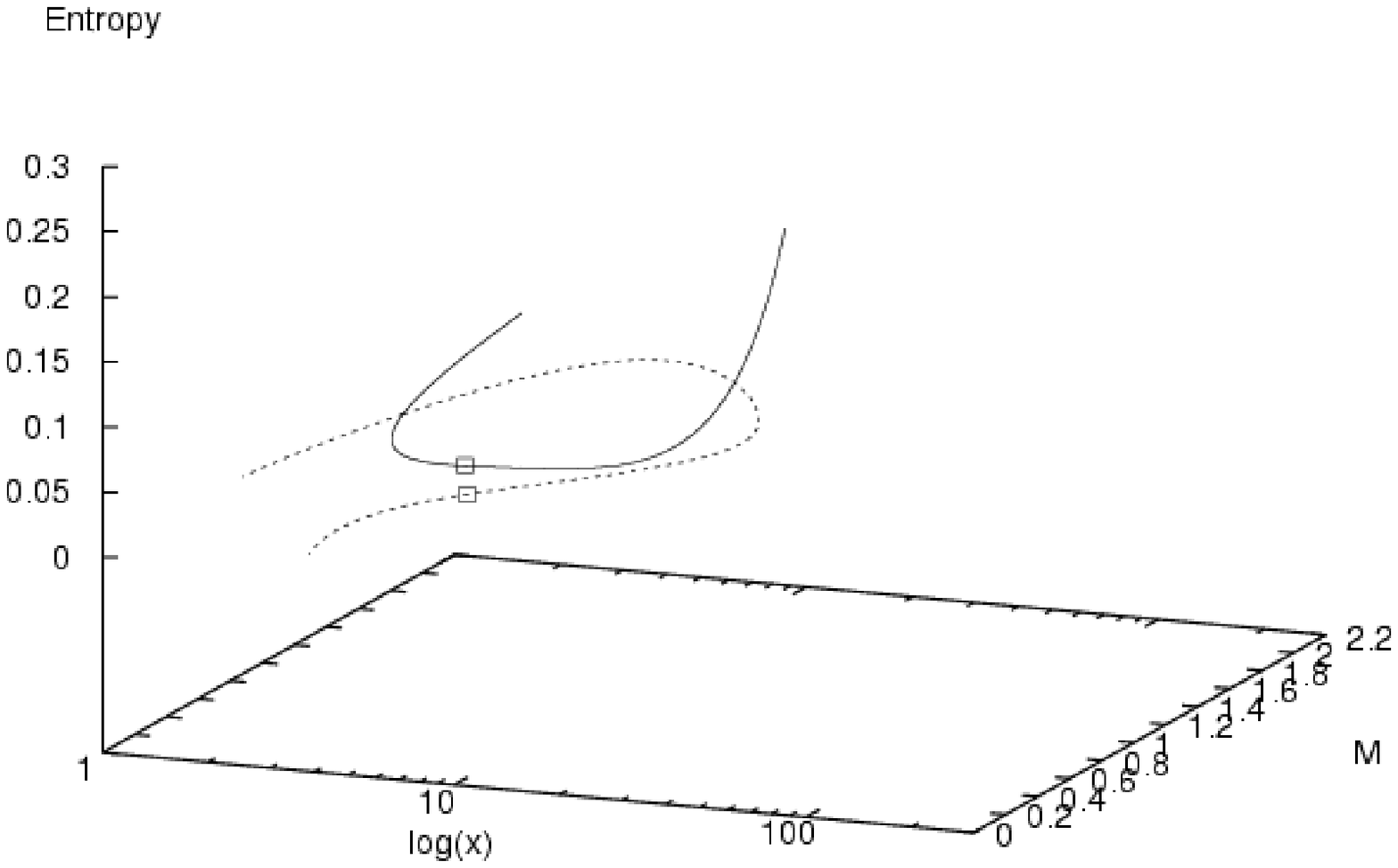,height=8truecm,width=8truecm}
}}
\noindent{\small {\bf Fig. 6:} 
Three-dimensional plot of the second panel ($\zeta=10$) of Fig. 5a
in which the specific entropy is also plotted as it varies along the flow.
The boxes represent the pseudo-intersection point of that panel at around $M\sim 0.5$.
The two branches have different entropies. }
\end{figure}

\section{Parameter Space Description}

It is useful to study the global behaviour of the accretion solutions. For this we 
classify the parameter space spanned by the specific angular momentum and energy at the
sonic point. Figure 7 shows the parameter spaces in which the solution passing 
through the inner sonic point contains a closed spiraling topology as in 
panel 1 of Fig. 5a. This means that whether the Rankine-Hugoniot relation is satisfied or not, a 
shock could form in this region. The shock will be stationary if the Rankine-Hugoniot 
relation is satisfied$^{14}$ and will be oscillating if the relation 
is not satisfied. The solid, dashed, dot-dashed and dot-long-dashed regions are for cooling 
parameter $1$, $10$, $30$, and $50$ respectively. As the cooling is increased, the region
shrinks and becomes smaller and smaller. This indicates that there exists a
critical cooling parameter, beyond which a flow will cease to have 
three sonic points. Other parameters used are $\alpha_\Pi=0.01$ and ${\dot m}=0.1$.
When the viscosity and cooling are reduced to zero, this region merges exactly to the corresponding region in C89.

\begin {figure}
\vbox{
\centerline{
\psfig{figure=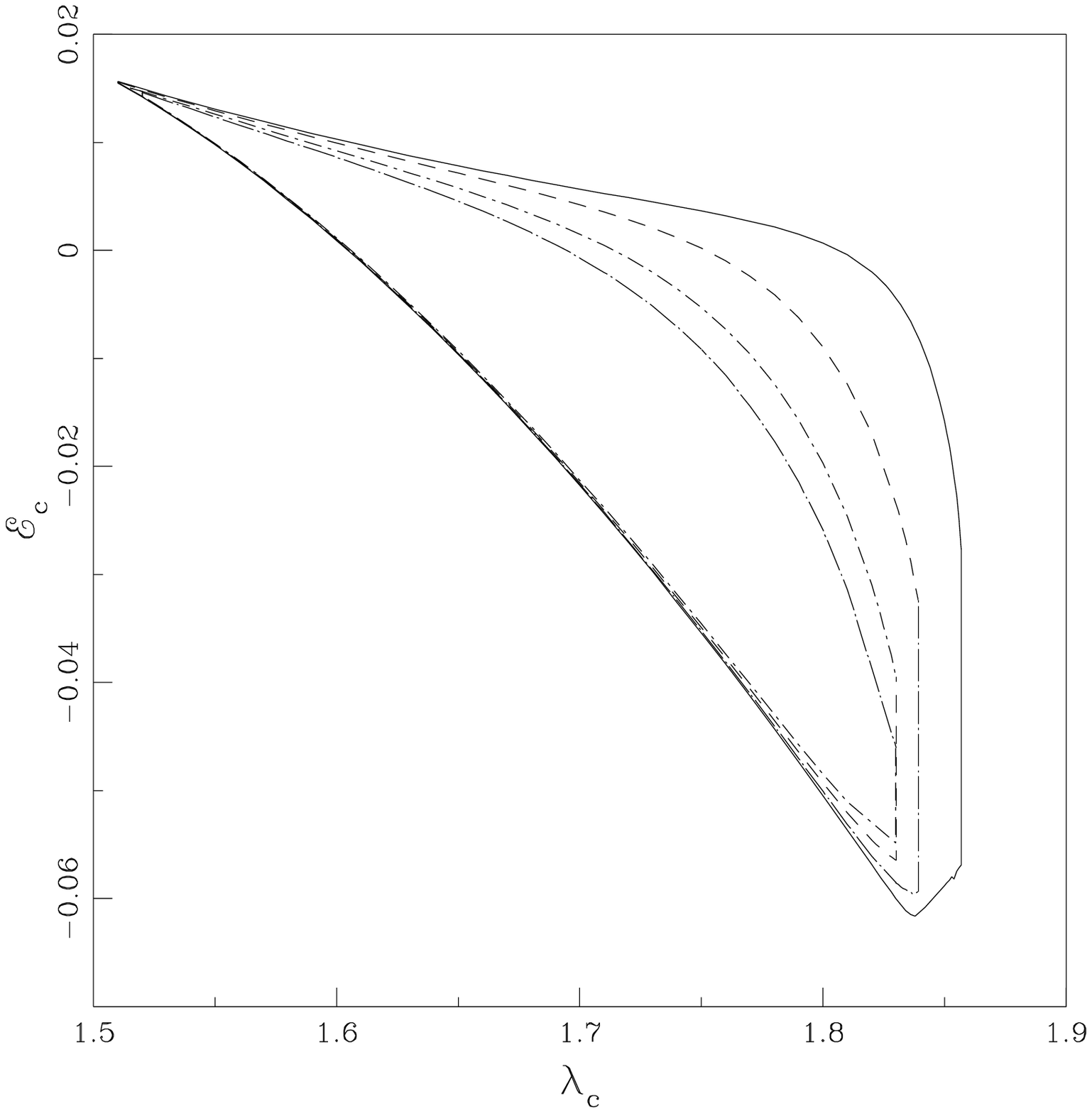,height=10truecm,width=10truecm}
}}
\noindent{\small {\bf Fig. 7:} 
Region of the parameter space in which the solution passing
through the inner sonic point contains a closed spiraling topology.
The solid, dashed, dot-dashed and dot-long-dashed regions are for cooling
parameter $1$, $10$, $30$, and $50$ respectively.  Other parameters are $\alpha_\Pi=0.01$ and ${\dot m}=0.1$.
}
\end{figure}

In Fig. 8(a-c), we present a few complete solutions which are drawn with the parameters at 
$\lambda_{c}=1.7, \ \alpha_\Pi=0.05, \ {\dot m}=0.2$ and $\zeta=5$ and only the inner sonic 
point is varied: (a) $x_{in}=2.545$, (b) $x_{in}=2.55$ and (c) $x_{in}=2.555$. The corresponding 
shock locations are (a) $x_{s}=48.199$, (b) $x_s=27.8854$ and (c) $x_s=18.6445$ respectively.
Vertical dashed lines show the shock transitions which connect two solutions, one passing through 
the inner sonic point and the other passing through the outer sonic point. The spiral loop through 
the inner sonic point rapidly shrinks with the increase in the sonic point location. The shock 
location also comes closer. This shows that even though the shock location may change by 
orders of magnitude, the inner sonic point virtually remains at the same place. 

\begin {figure}
\vbox{
\centerline{
\psfig{figure=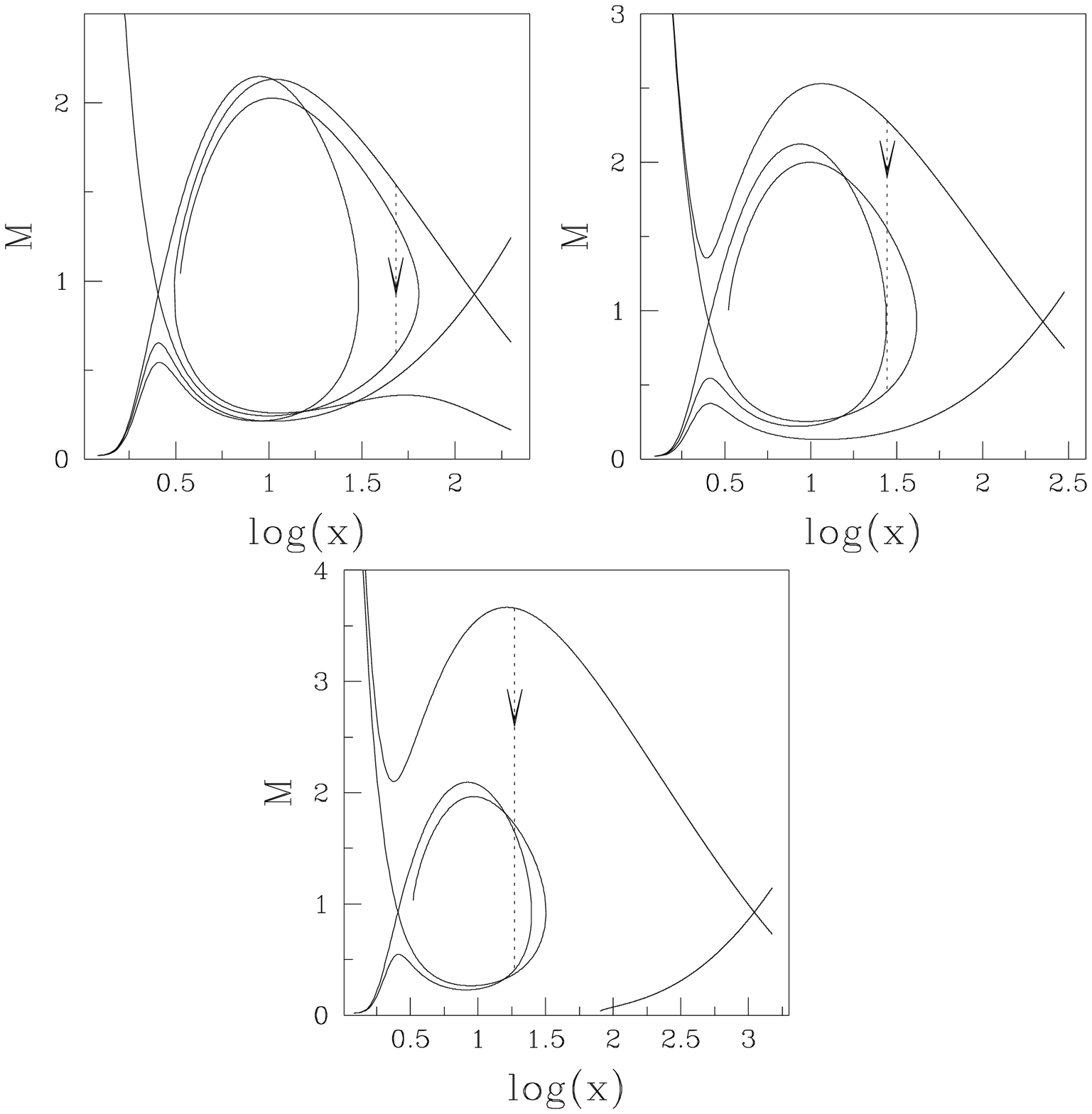,height=10truecm,width=10truecm}
}}
\noindent{\small {\bf Fig. 8(a-c):} 
A few complete solutions which are drawn with the parameters at
$\lambda_{c}=1.7, \ \alpha_\Pi=0.05, \ {\dot m}=0.2$ and $\zeta=5$ and only the inner sonic
point is varied: (a) $x_{in}=2.545$, (b) $x_{in}=2.55$ and (c) $x_{in}=2.555$. The corresponding
shock locations are (a) $x_{s}=48.199$, (b) $x_s=27.8854$ and (c) $x_s=18.6445$ respectively.
Vertical dashed lines show the shock transitions.
}
\end{figure}

It is instructive to know the sub-division of the parameter space in terms of the topologies of the
solutions. All the topologies seen in Paper I are also present in this case, but a new topologies occur
(see, Fig. 5b) when the cooling is especially strong. Figure 9 shows the sub-division of the 
parameter space which are marked and the corresponding topologies are shown in the bottom 
left. When the cooling is very strong, the curve ${\cal A}{\cal B}{\cal C}$ shows further sub-divisions
and a new topology shown in the box C13 occur. The number of loops in the inflow may increase
depending on the cooling. The dotted curves indicate that the Figure is drawn for 
different (high) cooling factor $\zeta$. The regions marked $S, \ OS, \ C11, \ C12, \ OAC, \ I*, \ O* $
produce topologies which produce standing shocks in accretion, oscillating shocks 
in accretion, region which produces one type of closed topology (clockwise
turn), region which produces the other type of closed topology (anti-clockwise  turn),
region which produces one open and the other closed topology, region which 
produces open solutions passing only through the outer saddle type sonic point
and the region which produces open solutions passing only through the inner saddle type sonic point
respectively. The region producing the new dotted topology C13 is very close
to the curve ${\cal A}{\cal B}{\cal C}$ and can be discernible only when cooling is strong.

\begin {figure}
\vbox{ 
\centerline{
\psfig{figure=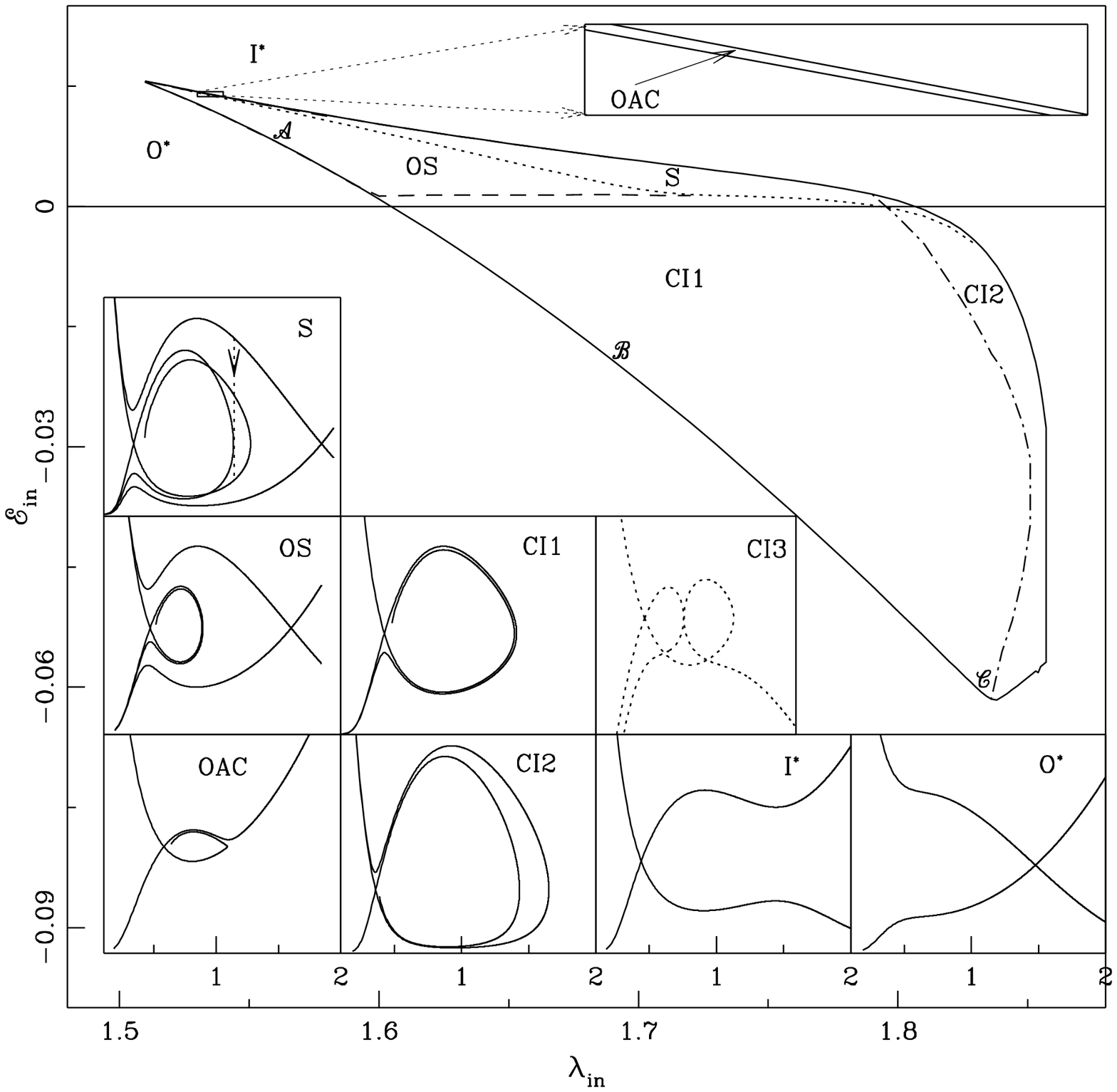,height=10truecm,width=8truecm}
}}
\noindent{\small {\bf Fig. 9:}
Division of the parameter space according to the solution topologies shown in the 
inset. Details are in the text.
}
\end{figure}
 
It would be of interest, to concentrate on the modification of the 
parameter space for shock formation when cooling is enhanced. In Figs. 10 (a-b)
this is shown. In Fig. 10a, $\alpha_\Pi=0.01$ is chosen and the cooling parameters are $\zeta=0.01$, (dot-dashed)
$0.1$ (long-dashed) and $1$ (solid). We note that the region of the parameter space shifts to 
include negative energy regions as well (For instance, for $\alpha_\Pi=0$ and $\zeta=0$, the 
parameter space contains only positive energy.). In Fig. 10b, viscosity parameter is increased
to $\alpha_\Pi=0.05$. This causes a shrinkage in the parameter space. The cooling parameters
are $\zeta=1$ (dot-dashed), $5$ (long-dashed) and $10$ (solid) respectively. In this case, the
parameter space shrinks drastically when cooling is enhanced.

\begin {figure}
\vbox{ 
\centerline{
\psfig{figure=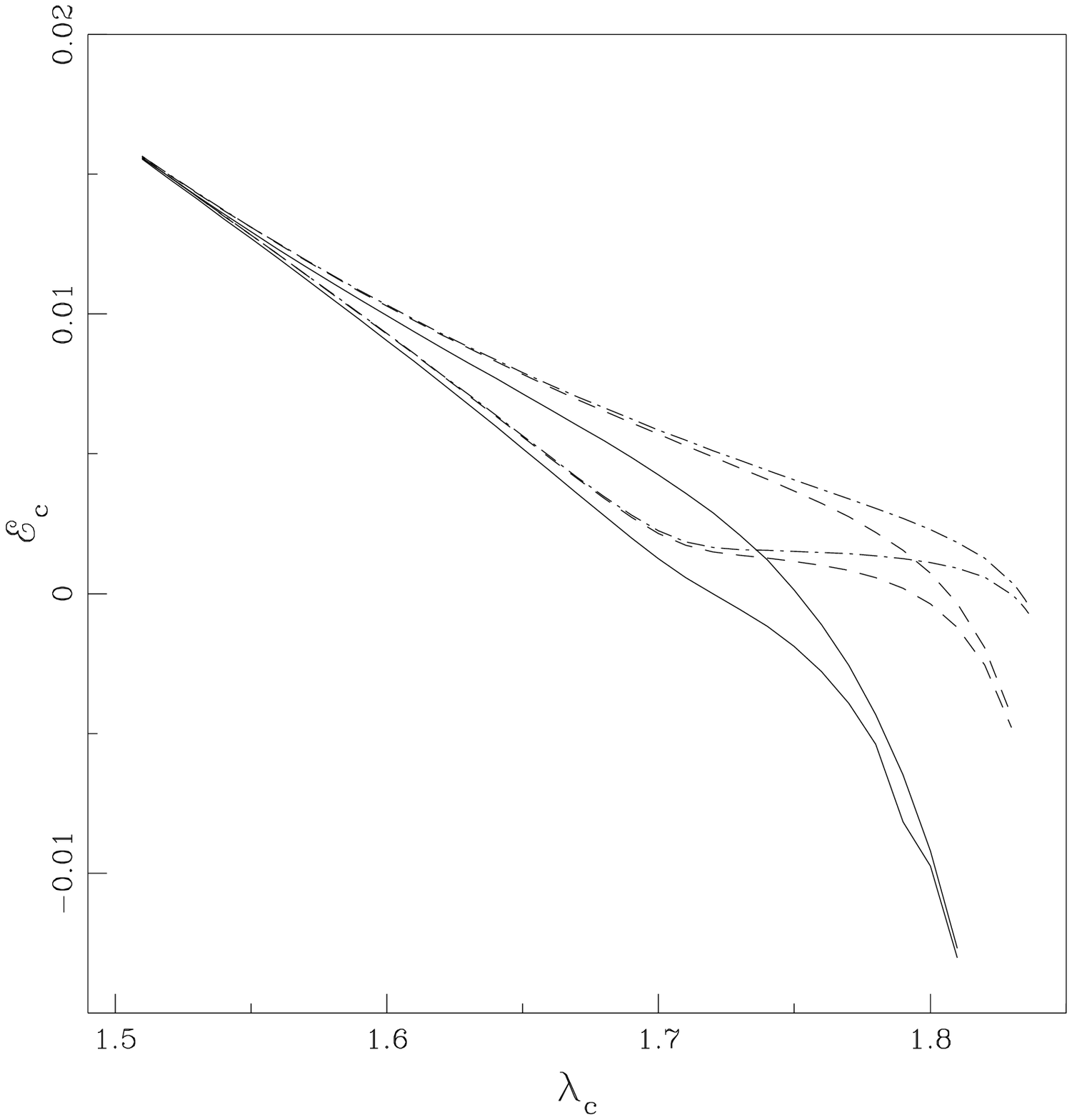,height=10truecm,width=8truecm}
\psfig{figure=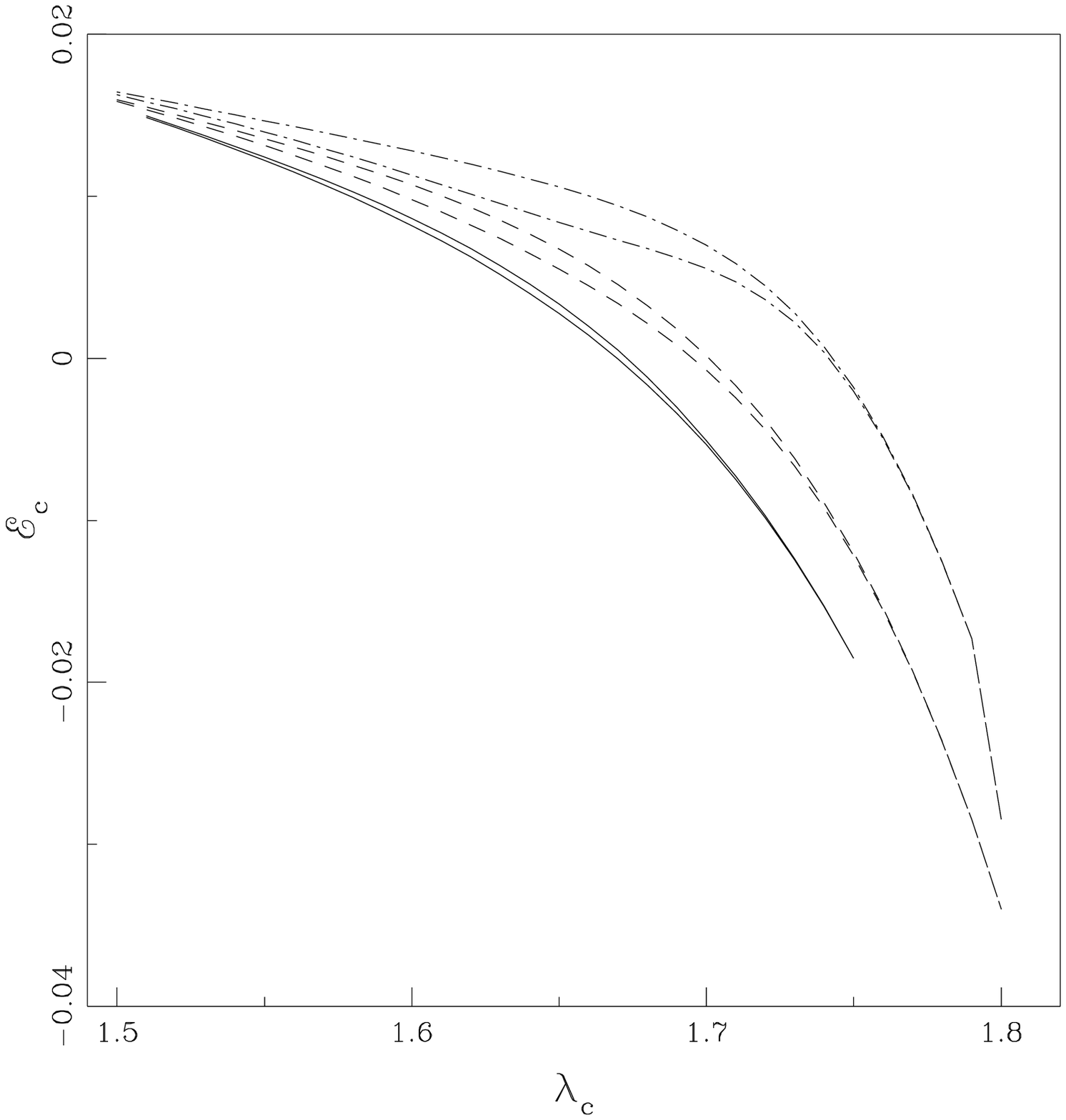,height=10truecm,width=8truecm}
}}
\noindent{\small {\bf Fig. 10(a-b):}
Modification of the parameter space for shock formation when cooling is varied.
(a) $\alpha_\Pi=0.01$ is chosen and $\zeta=0.01$, (dot-dashed) $0.1$ (long-dashed) and $1$ (solid) (left panel).
(b) $\alpha_\Pi=0.05$ is chosen and $\zeta=1$ (dot-dashed), $5$ (long-dashed) and $10$ (solid) (right panel).
}
\end{figure}

\section{Sonic Points for Super-Keplerian Flows?}

It is generally accepted that at the sonic points, the flow must be sub-Keplerian (C90b). However,
if the cooling is very efficient, this requirement may be violated. In Fig. 11, 
we show that when $\zeta$ is very high, of the order of a few hundreds, the sonic point 
may have angular momentum {\it above} the Keplerian distribution (dotted curve). 
Particularly important is that this is possible if the sonic point is located 
near the black hole horizon. The implication of this is not obvious. Does it mean that 
very cold flow {\it can} be Keplerian or super-Keplerian throughout its 
journey? If so, can it spin up the black hole faster that what is presumed so far? 
This point is to be addressed in future. 

\begin {figure}
\vbox{
\centerline{
\psfig{figure=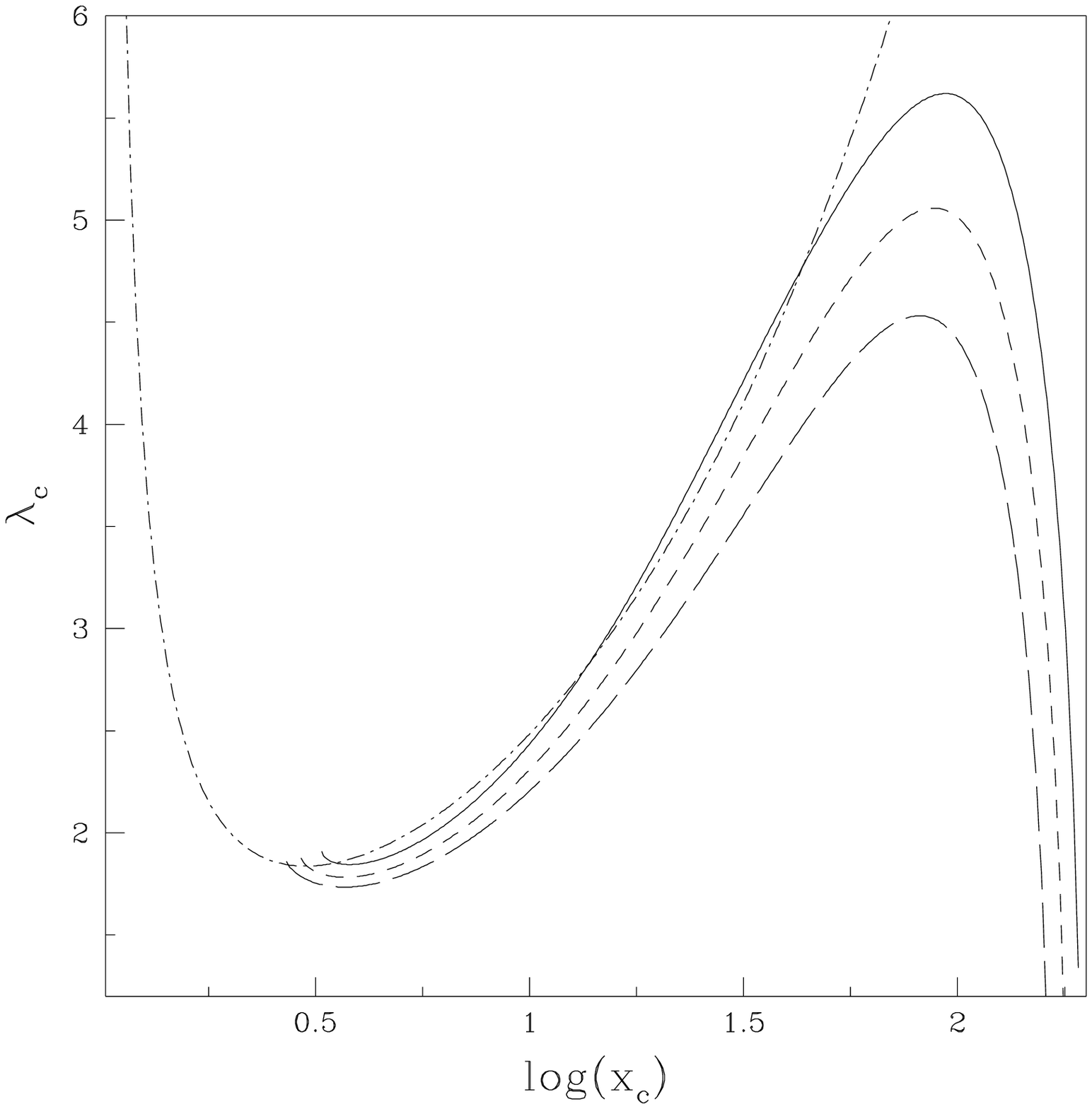,height=10truecm,width=10truecm}
}}
\noindent{\small {\bf Fig. 11:}
Example of parameters which produce transonic solutions with super-Keplerian
angular momentum. The dot-dashed curve is the Keplerian distribution. Solid
curves, from the bottom to the top, are for $\zeta=400, 500$ and $600$ respectively. }
\end {figure}

\section {Critical Cooling and Sub-division of the Parameter Space}

We have already indicated that cooling and heating have opposite effects in
deciding the solution topologies, but one does not {\it exactly} cancel 
the other effect. When the cooling is enhanced for a given viscosity parameter,
the possibility of shock formation is eventually reduced. This is shown in Fig. 12.
Here the critical cooling parameter $\zeta=\zeta_{cri}$ is plotted against the 
specific angular momentum for two different viscosity parameters. Solid curve is
for $\alpha_\Pi=0.01$ and the dashed curve is for $\alpha_\Pi=0.05$. All possible
inner sonic points are considered. The region below the curve contains 
topologies which are closed and therefore standing or oscillating shocks
could be possible while the region above the curve allows solutions with open topologies.
We note that for smaller $\lambda$, the critical cooling factor is small
which for larger $\lambda$. This is expected since the possibility of shock formation is
enhanced with $\lambda$ in general. When $\alpha$ is higher $\zeta_{cri}$ is lower. This 
indicates that general reduction of the parameter space due to higher viscosity$^1$. 
is not totally compensated for by cooling effects.

\begin {figure}
\vbox{
\centerline{
\psfig{figure=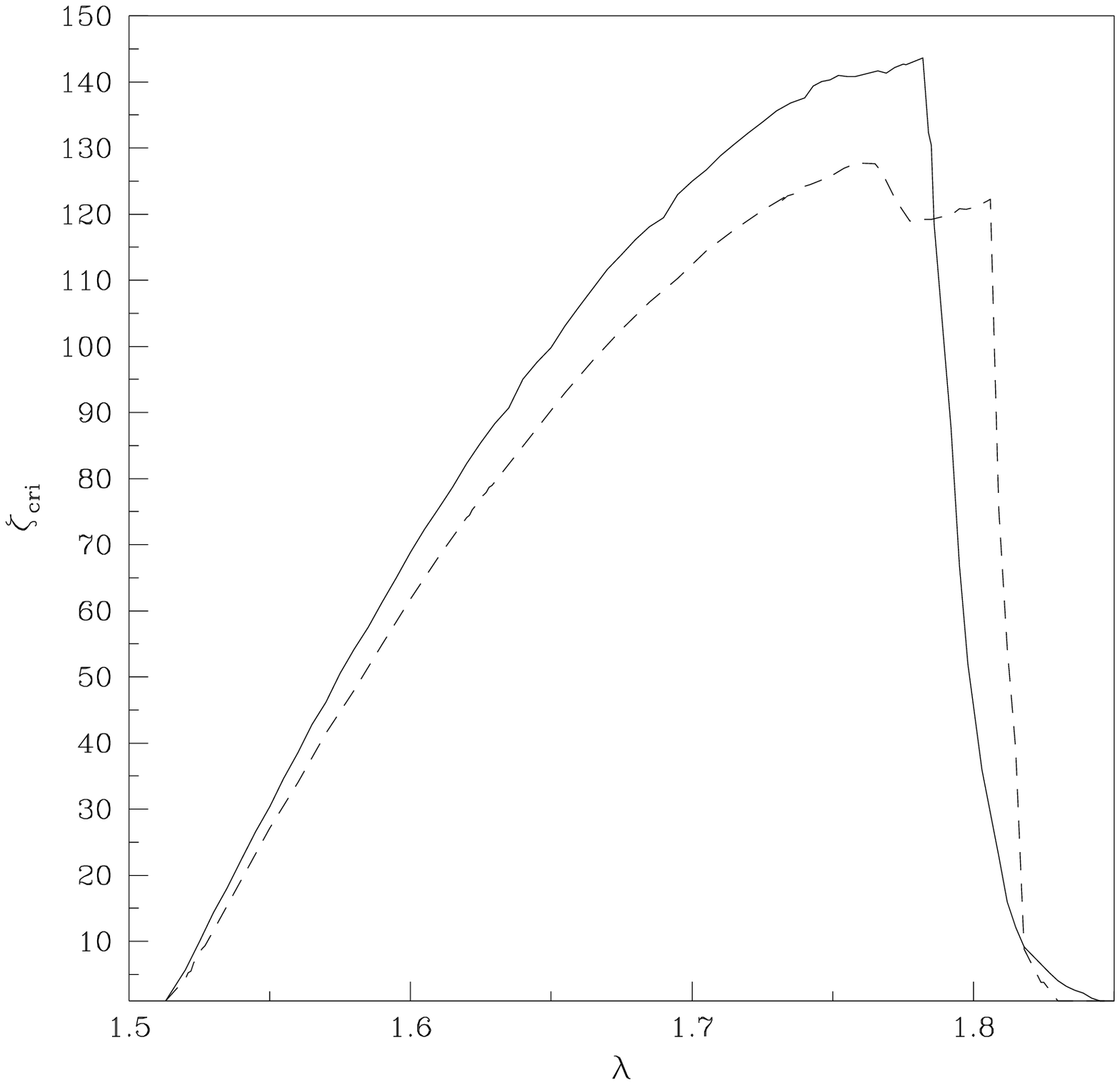,height=10truecm,width=10truecm}
}}
\noindent{\small {\bf Fig. 12:}
The variation of the critical cooling parameter as a function of the specific angular momentum
at the inner sonic point for two different viscosity parameters. Solid curve is for $\alpha_\Pi=0.01$ and the dashed curve is for $\alpha_\Pi=0.05$. The region below the curve contains 
topologies which are closed and therefore standing or oscillating shocks
could be possible while the region above the curve allows solutions with open topologies.
}
\end {figure}

\section{Concluding Remarks}

In this paper, we studied the dissipative accretion flow in presence of
viscous heating and bremsstrahlung cooling processes.
Viscosity tends to heat the flow, thereby reducing the
Mach number. Cooling, on the other hand, increases the Mach number. Thus,
formation of shock, which involves a supersonic to subsonic transition is affected
by heating and cooling. We classified the parameter space in terms of whether 
three sonic points or  shocks can form or not. We discovered a completely new topology
in which matter, coming from a large distance, is connected to the black hole horizon
as a normal solution, but it has multiple valued Mach number solution.
We find that cooling can `undo' the effect of heating
on topological properties, only to certain extent. If the viscosity is 
high enough, then no matter how much cooling is used, the parameter space shrinks.
We have also found that for a given set of flow parameters, there is always  
a critical cooling factor which separates the parameter space into closed regions, 
one with a closed and the other with an open topology.

It is generally believed that only the sub-Keplerian flows can pass through 
the sonic points. However, we find that when the flow is very strongly cooled,
even super-Keplerian flows can also pass through the inner sonic point. This is 
a new result and may be significant in evolution of the spin of the accretion black holes.

This work is partly supported by a project (Grant No. SP/S2/K-15/2001)
funded by Department of Science and Technology (DST).

{}

\end{document}